\documentclass[%
 reprint,
superscriptaddress,
 amsmath,amssymb,
 aps,
 pra,
]{revtex4-2}

\usepackage{graphicx}% Include figure files
\usepackage{siunitx}
\usepackage{pgfplots}% Include matplotlib pgf plots
\pgfplotsset{compat=1.18}

\usepackage{dcolumn}% Align table columns on decimal point
\usepackage{bm}% bold math
\usepackage{hyperref}% add hypertext capabilities
\hypersetup{
	colorlinks,
	linkcolor={black},
	citecolor={black},
	urlcolor={blue}
}
\usepackage{tikz}
\usetikzlibrary{decorations.pathreplacing}
\usetikzlibrary{arrows.meta}
\usetikzlibrary{decorations.pathmorphing}
\usetikzlibrary{positioning}
\usetikzlibrary{calc}
\usetikzlibrary{patterns}

\usepackage{adjustbox}
\usepackage{makecell}

\usepackage{xfrac}

\usepackage{booktabs}

\usepackage[T1]{fontenc}   % better glyph coverage & font selection
\usepackage{lmodern}        % scalable CM (Latin Modern)

\usepackage{xspace}

\newcommand{\Esca}{\mathcal{E}}

\newcommand{\cm}{\,\mathrm{cm}}

\newcommand{\CENTREX}{CeNTREX\xspace}

\newcommand{\RtwoFfour}{$R(2)\,\tilde{F}^\prime_1=7/2\,F^\prime=4$\xspace}

\newcommand{\PtwoFone}{$P(2)\,\tilde{F}^\prime_1=3/2\,F^\prime=1$\xspace}
\newcommand{\PtwoFtwo}{$P(2)\,\tilde{F}^\prime_1=3/2\,F^\prime=2$\xspace}

\newcommand{\ket}[1]{\left|#1\right\rangle}

\begin{document}

\preprint{APS/PRA}

\title{\texorpdfstring{Electrostatic quadrupole lens for focusing a cryogenic $^{205}$TlF molecular beam}{Electrostatic quadrupole lens for focusing a cryogenic 205TlF molecular beam}}% Force line breaks with \\

\author{Pengyu Zhou}
\email{pz2297@columbia.edu}
\affiliation{Department of Physics, Columbia University, New York, New York 10027, USA}

\author{Olivier Grasdijk}
% \email{jgrasdijk@argonne.gov}
\affiliation{Physics Division, Argonne National Laboratory, Argonne, Illinois 60439, USA}

\author{David DeMille}
\affiliation{Physics Division, Argonne National Laboratory, Argonne, Illinois 60439, USA}
\affiliation{Department of Physics and Astronomy, Johns Hopkins University, Baltimore, Maryland 21218, USA}

\author{Jakob Kastelic}
\affiliation{Department of Physics, Yale University, New Haven, Connecticut 06511, USA}

\author{David Kawall}
\affiliation{Department of Physics, University of Massachusetts Amherst, Amherst, Massachusetts 01003, USA}

\author{Jianhui Li}
\affiliation{Department of Physics, Columbia University, New York, New York 10027, USA}

\author{Oskari Timgren}
\affiliation{Department of Physics, Yale University, New Haven, Connecticut 06511, USA}

\author{Konrad Wenz}
\affiliation{Department of Physics, Columbia University, New York, New York 10027, USA}

\author{Yuanhang Yang}
\affiliation{Department of Physics, University of Chicago, Chicago, Illinois 60637, USA}

\author{Tanya Zelevinsky}
\affiliation{Department of Physics, Columbia University, New York, New York 10027, USA}

\collaboration{CeNTREX Collaboration}
\noaffiliation

\date{\today}% It is always \today, today,

\begin{abstract}
Precision measurements with molecules require molecular beams with high flux and narrow divergence. Here we present the design, implementation, and characterization of an electrostatic quadrupole lens (EQL) for focusing a cryogenic beam of $^{205}$TlF molecules. The EQL consists of four electrodes operated at spatially alternating potentials up to $\pm 30$~kV. This configuration generates a transverse restoring force that focuses molecules prepared in selected manifold of hyperfine states. 
% the $\ket{J=2,m_J=0}$ state manifold. 
At the optimal applied voltage, the EQL increases the detected molecular signal by a factor of $14.4(4)$ at the current detection position. The measured lens gain, transverse Doppler spectra, and transverse spatial profiles are all reproduced by trajectory simulations using independently measured beam parameters as input.
%including the source position distribution of the beam.
Based on this agreement, we project a gain of 16.5(8) for the full $7.9~\mathrm{m}$ \CENTREX{} beamline. These results establish the EQL as a key component for increasing sensitivity in precision experiments using cryogenic beams of polar molecules.
% \begin{description}
% \todo{Secondary publications and information retrieval purposes.}
% You may use the \texttt{description} environment to structure your abstract;
% use the optional argument of the \verb+\item+ command to give the category of each item.
% \end{description}
\end{abstract}

 
\maketitle

%\tableofcontents

\section{Introduction}
\label{sec:introduction}

Molecular beams are a standard platform for precision measurements. The statistical sensitivity of these measurements generally improves with the number of detected molecules in the relevant quantum state.
However, geometric constraints between the beam source and detection can significantly reduce the detected molecule number. This is especially pronounced for cryogenic buffer-gas beams that offer low forward velocities but present substantial transverse divergence \cite{hutzler2012buffer}. Transverse collimation and focusing therefore become particularly useful for increasing the detected molecular flux and the statistical sensitivity.

Electrostatic multipole lenses are commonly used to control the transverse motion of polar molecules. By exploiting the Stark shifts of selected states, electrostatic multipole fields can produce state-dependent and position-dependent transverse forces, enabling collimation, focusing, and state selection in molecular beams~\cite{scoles1992atomic,etde_20124735,ramsey1956molecular}. These techniques have long been used in molecular beam physics and precision measurements, including earlier thallium monofluoride (TlF) Schiff-moment searches as well as recent experiments with ThO and BaF molecules~\cite{cho1991search,PhysRevA.29.425,PhysRevLett.63.2559,PhysRevLett.59.991,Wu_2022,Touwen_2024,TownesGordonPR55_Maser}. These demonstrations established electrostatic focusing as a powerful technique with performance that depends strongly on the molecular beam source properties and internal state preparation.

In \CENTREX{} (the Cold Molecule Nuclear Time-Reversal EXperiment), electrically polarized TlF molecular beams are used to search for the nuclear Schiff moment of $^{205}$Tl, improving constraints on hadronic time-reversal violation~\cite{Grasdijk_2021}. In the experiment's $7.9~\rm m$ long beamline, transverse expansion of the molecular beam can strongly limit the number of molecules reaching the detection region. Increasing the detected molecular flux therefore requires sufficient accumulation in the desired state as well as efficient transverse focusing and collimation through the beamline. Considerable progress has been made in population transfer and state preparation via rotational-hyperfine cooling~\cite{grasdijk_rotational_cooling_2025} and microwave-driven adiabatic passage~\cite{CeNTREX_state_preparation_a}. Electrostatic quadrupole lensing can address collimation and focusing while preserving compatibility with the state-preparation sequence.

Here we implement and characterize an electrostatic quadrupole lens for \CENTREX{}. We first characterize the molecular source position distribution, which provides an important input for quantitative modeling of the lens performance. We then measure the increase in detected molecule number produced by the lens and characterize the effect of the lens on velocity and position distributions of molecules. This work extends previous TlF electrostatic-focusing measurements in several important ways. First, the lens is applied to a cryogenic buffer-gas TlF beam, whose relatively slow forward velocity and narrow velocity spread make transverse focusing particularly effective. Second, the focusing is performed in a selected internal state, enabling full control over the molecules' internal and external degrees of freedom. Third, the lens operates at electric fields up to $\sim30~\mathrm{kV/cm}$,
%and focuses molecules in the $\ket{J=2,m_J=0}$ manifold, where $J,m_J$ are rotational quantum numbers,
enabling strong focusing. Near the beam axis, the electric field of the quadrupole lens creates a harmonic transverse potential; hence, the device functions analogously to a thick optical lens~\cite{berg1965determination,cho1991tight}.

\section{Background}
\label{sec:background}
In this section, we briefly present the theoretical background for the electrostatic quadrupole lens principle and the properties of TlF molecules in electric fields.
\subsection{Lens Model}
\label{subs:lens_model}
\paragraph*{Lens field and quadratic Stark potential.}
The quadrupole configuration of ideal electrodes held at spatially alternating electric potentials $\pm V$ generates an electric field of magnitude
\begin{equation}
\label{eq:equad}
    \bigl|\Esca(r)\bigr| = \frac{2 V r}{R^2},
\end{equation}
where $2R$ is the lens bore diameter and $r$ is the distance from the beam axis. For a molecular state with quadratic Stark shift, the quadrupole field produces a harmonic radial potential,
\begin{equation}
\label{eq:quad_stark_potential}
    \Delta E_S^{(2)}(r)
    =
    C\bigl|\Esca(r)\bigr|^2
    =
    4C\frac{V^2}{R^4}r^2 ,
\end{equation}
where $\Delta E_{S}^{(2)}$ denotes the Stark shift evaluated in the
quadratic approximation, and $C$ is a state-dependent coefficient. For a linear rigid rotor in the weak-field limit, second-order perturbation theory gives the standard result~\cite{browncarrington}; for the lensing state $\ket{J=2,m_J=0}$ ($J$ and $m_J$ are the rotational quantum number and its projection), this reduces to
\begin{equation}
    C=\frac{\mu_e^2}{42hB},
    \label{eq:quad_stark_potential_J2mJ0}
\end{equation}
where $B$ is the rotational constant, $\mu_e$ is the body-fixed electric dipole moment, and $h$ is Planck's constant.  In the case considered here, $\mu_e=4.23$~Debye and $B=6.67$~GHz.

For TlF, this weak-field quadratic description is appropriate because the $X\,^1\Sigma^+$ ground state lacks opposite-parity near-degenerate levels and the $\ket{J=2,m_J=0}$ Stark shift remains very nearly quadratic in fields up to $\Esca\approx20~\rm kV/cm$. For the maximum electrode voltage used here, $V=30~\mathrm{kV}$ and the electric field inside the lens bore is below $30~\mathrm{kV/cm}$. Thus the analytic quadratic model provides a good description for near-axis trajectories and is a useful guide to the lens behavior.

\paragraph*{Focusing strength and effective focal length.}
The harmonic Stark potential corresponds to a radial restoring force
\begin{equation}
    F_r
    =
    -\frac{\partial \Delta E_S^{(2)}}{\partial r}
    =
    -8C\frac{V^2}{R^4}r .
\end{equation}
For a molecule of mass $m$ with approximately constant longitudinal velocity $v_Z$, using $Z=v_Z t$ (where $t$ is the time of traversal) gives
\begin{equation}
    \frac{d^2r}{dZ^2}+p^2r=0,
\end{equation}
where
\begin{equation}
    p =
    \left(
    \frac{8CV^2}{R^4mv_Z^2}
    \right)^{1/2}.
    \label{eq:focusing_parameter}
\end{equation}
Thus the EQL acts as a finite-length focusing element with a focusing strength (the spatial wavenumber $p$) proportional to $\left|V\right|/v_Z$. In the
ideal harmonic approximation, the EQL can be described using the ray transfer matrix formalism which relates the transverse position and angle of a molecule at the lens entrance and exit. This model treats an EQL of length $l$ as a thick optical lens whose entrance and exit planes are separated by $l$, with an effective focal length~\cite{berg1965determination,cho1991tight}
% In the ideal harmonic approximation, a lens of length $l$ is described by the transfer matrix\cite{berg1965determination,cho1991tight}
% \begin{equation}
%     M(l)=
%     \begin{pmatrix}
%         \cos(pl) & p^{-1}\sin(pl) \\
%         -p\sin(pl) & \cos(pl)
%     \end{pmatrix}.
% \end{equation}
% The effective focal length follows from the lower-left matrix element,
\begin{equation}
    f_{\rm eff}=\frac{1}{p\sin(pl)}.
    \label{eq:effective_focal_length}
\end{equation}

Because $p\propto |V|/v_Z$, the focal length depends on longitudinal velocity. The finite velocity spread of the molecular beam therefore produces a spread of focal lengths, analogous to chromatic aberration in an optical lens.

\subsection{TlF in Electric Fields}
\label{sec:state_labeling}

In the electronic and vibrational ground state $X\,^1\Sigma^+(v=0)$ of TlF, we typically denote field-free states under the coupled basis $\ket{J,F_1,F,m_F}$~\cite{Grasdijk_2021} where $J$ is the total angular momentum excluding the nuclear spins; $\mathbf{F}_1=\mathbf{J}+\mathbf{I}_1$ with $I_1=1/2$ for $^{205}$Tl; $\mathbf{F}=\mathbf{F}_1+\mathbf{I}_2$ with $I_2=1/2$ for $^{19}$F; and $m_F$ is the projection of $\mathbf{F}$ along the quantization axis. At intermediate fields, $J$ and $m_J$ are approximately good quantum numbers. 

%At zero electric field, the coupled basis are the eigenstates.
At higher fields ($\Esca \gtrsim 10~\mathrm{kV/cm}$), the Stark interaction becomes comparable to the rotational spacing, and greatly exceeds the hyperfine couplings, mixing adjacent $J$ levels so that $J$ ceases to be a good quantum number. Nevertheless, the $J=2,m_J=0$ Stark shift stays close to quadratic up to $\Esca \approx 20~\mathrm{kV/cm}$ before higher-order corrections become significant (see Fig. \ref{fig:quadrupole_lens} (a)).
%the Stark interaction dominates over both rotational and hyperfine couplings and $J$ is no longer a good quantum number.

In the excited electronic state $B^3\Pi_1(v=0)$ used for detection, $J$ and $F_1$ are only approximate quantum numbers because of strong hyperfine mixing~\cite{norrgard2017hyperfine,meijer2020lambda} and the eigenstates are denoted by $\ket{\tilde{J}, \tilde{F}_1, F, m_F}$.

\section{Experiment Overview}
\label{sec:experimental_overview}

\subsection{Beamline and State Preparation}
\label{sec:beamline_state_prep}

\begin{figure*}
	\includegraphics[width=\linewidth]{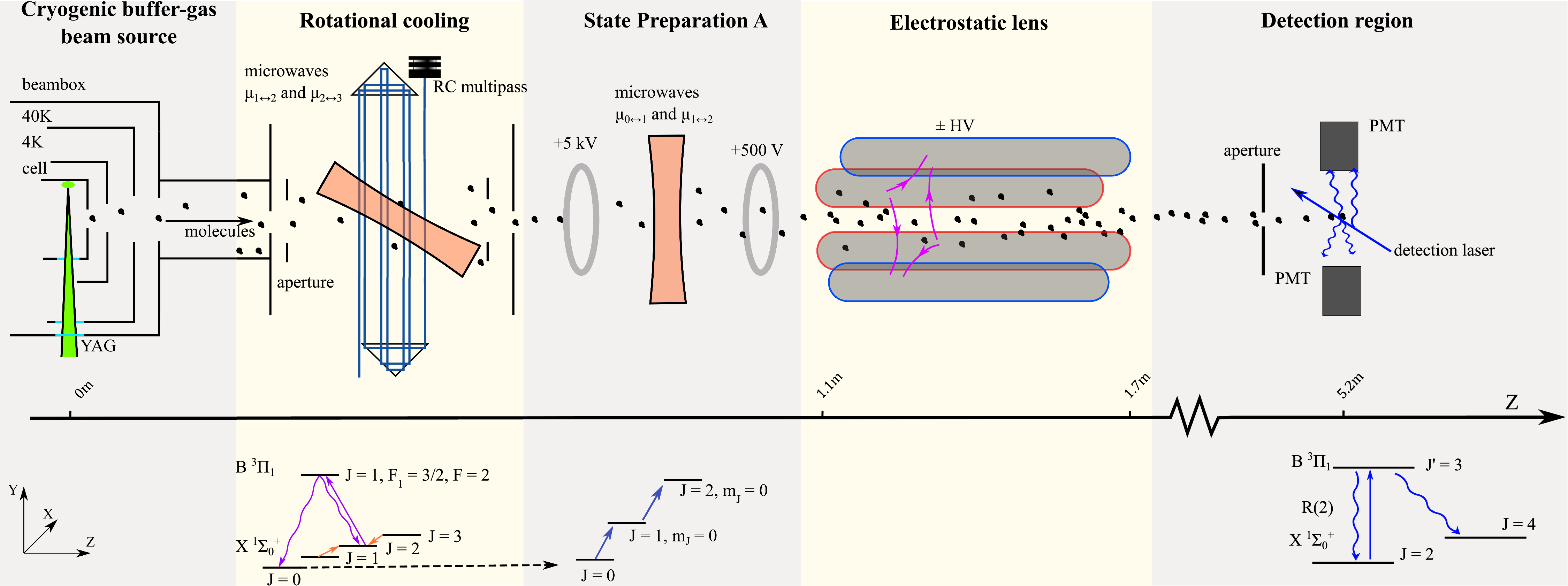}
	\caption{Schematic of the experimental apparatus. TlF molecules emerge from the cryogenic buffer-gas beam source and pass through the Rotational Cooling (RC) region, the State Preparation A (SPA) region, and the electrostatic quadrupole lens before entering the laser-induced fluorescence (LIF) detection region. The electric field lines produced by the lens are shown as purple arrows.  The molecular beam travels along the $Z$ direction; the other spatial axes are shown as well.
    %the vertical direction as $Y$, and the remaining transverse direction as $X$.
    Beam aperture positions are indicated.
    %The bottom of the plot indicates
    Molecular state evolution is illustrated in the lower part of the figure.}
	\label{fig:experiment_schematic} 
\end{figure*}
A schematic of the apparatus is shown in Fig.~\ref{fig:experiment_schematic}. For each molecular pulse, the cryogenic TlF beam has an approximately Gaussian forward velocity distribution, with a mean $\bar{v}_Z\approx195~\rm m/s$ and standard deviation $\sigma_{v_Z}\approx17~\rm m/s$.
%where $Z$ denotes the beam axis direction.
The pulse-to-pulse fluctuation of the mean velocity is $\sigma_{\bar{v}_Z}\approx5\rm m/s$,
which appears as a slow drift as the ablation target degrades. After exiting the source, the molecules enter the rotational cooling (RC) chamber, located $\sim40~\mathrm{cm}$ downstream from the source chamber and $\sim60~\mathrm{cm}$ from the cell exit. Here the molecules interact simultaneously with a multi-pass laser beam and two focused, free-space microwave beams that transfer population from the $J=1,2,3$ rotational levels into $J=0$~\cite{grasdijk_rotational_cooling_2025}. After RC, the molecules enter the State Preparation A (SPA) chamber, where a spatially varying electric field and two focused microwave beams drive successive adiabatic-passage transitions into the $\ket{J=2,m_J=0}$ manifold of hyperfine substates used for lensing~\cite{CeNTREX_state_preparation_a}. The prepared molecules then pass through the electrostatic quadrupole lens before reaching the laser-induced fluorescence (LIF) detection region.

\subsection{Electrostatic Quadrupole Lens}
\label{sec:eql}

\begin{figure}
  \centering
  \begin{minipage}{0.4\textwidth} % resize the whole figure here
    \begin{tikzpicture}
      % Place the image
      \node[anchor=south west, inner sep=0] (img) at (0,0)
        {\includegraphics[width=\linewidth]{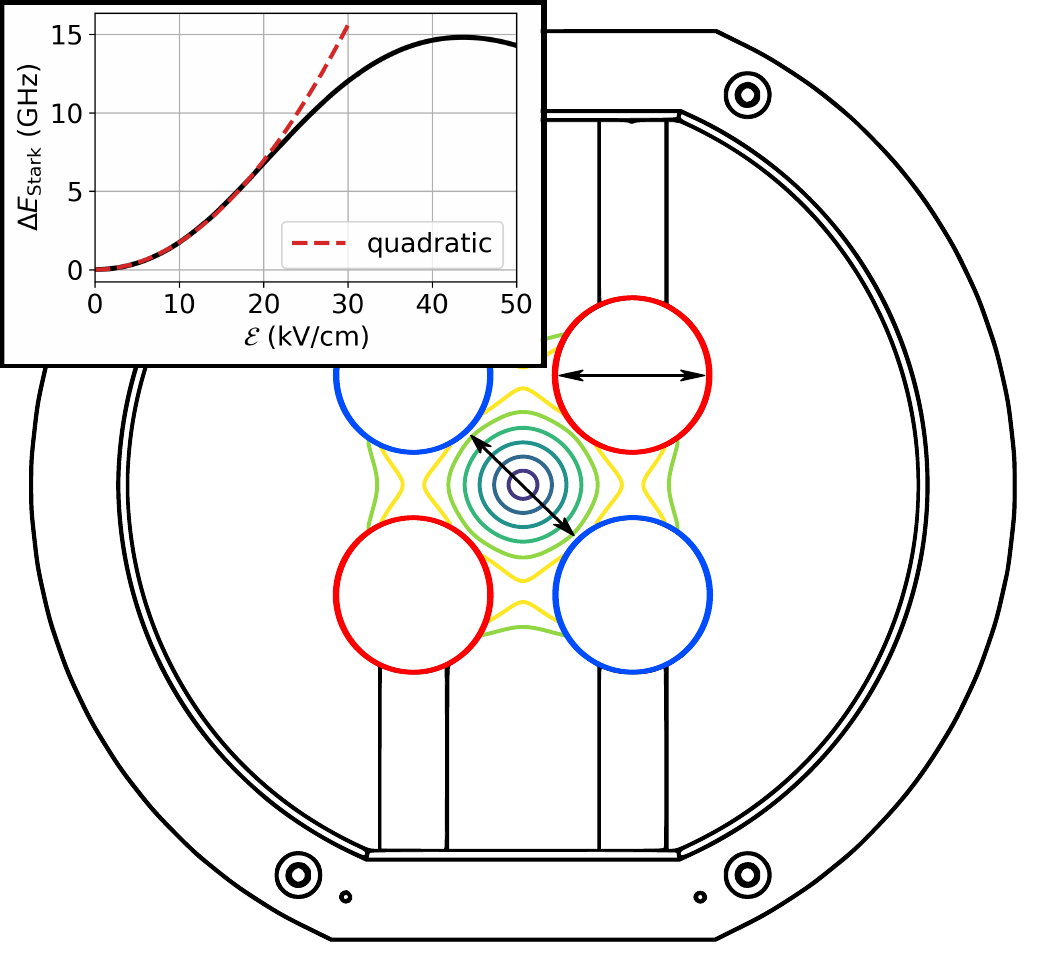}};

      % Normalized axes: origin at SW, x = SE-SW, y = NW-SW
      \begin{scope}[
        shift={(img.south west)},
        x={($ (img.south east) - (img.south west) $)},
        y={($ (img.north west) - (img.south west) $)}
      ]
        % Now (0,0)=bottom-left, (1,1)=top-right, so center is (0.5,0.5)
        \node at (0.640,0.500) {\small \colorbox{white}{$2R=1.75$"}};
        \node at (0.604,0.640) {\small $2R$};
        \node at (0.604,0.580) {\small $-V$};
        \node at (0.604,0.385) {\small $+V$};
        \node at (0.395,0.385) {\small $-V$};
        \node at (0.395,0.600) {\small $+V$};
        \node at (0.040,0.660) {\textbf{a)}};
        \node at (0.060,0.060) {\textbf{b)}};
      \end{scope}
    \end{tikzpicture}
  \end{minipage}
  \caption{(a) Stark shift of the %$J=2,\ m_J=0$
  $\ket{J=2,m_J=0}$ hyperfine manifold of states. The red dashed line indicates the quadratic Stark shift given by Eqs.~(\ref{eq:quad_stark_potential}--\ref{eq:quad_stark_potential_J2mJ0}) at $\mathcal{E}\lesssim20~\rm kV/cm$.  
  %where the Stark shift is approximately quadratic up to $20\rm kV$.
  (b) Front view of the electrostatic quadrupole lens (EQL). Colored curves are equipotential
  surfaces. The electrodes have length $l=60\cm$, the bore diameter and the rod diameter is $2R=1.75~\rm in$, and the applied potential can be as high as $V=30$ kV.
  %$\pm V$ up to $\pm30\,$kV.
  The electrode support structure is mounted on translation stages (not shown) that allow for
  alignment of the lens under vacuum. (Reproduced from Ref.~\cite{Grasdijk_2021}.)
  \label{fig:quadrupole_lens}}
\end{figure}
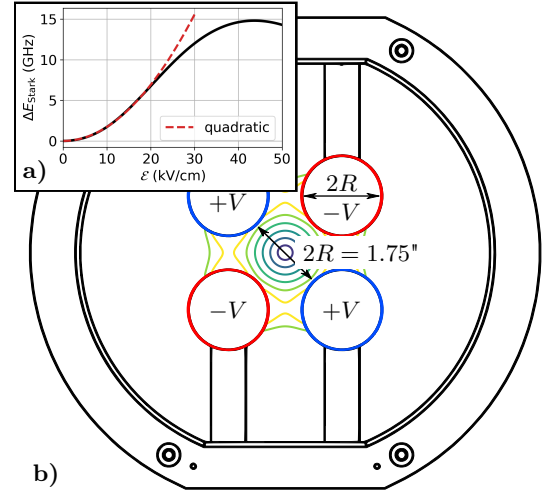
The electrostatic quadrupole lens (EQL) is located immediately downstream of the SPA chamber and focuses molecules in the $\ket{J=2,m_J=0}$ manifold populated by the adiabatic-passage sequence. The lens assembly consists of four cylindrical stainless-steel electrodes arranged symmetrically about the beam axis, as shown in Fig.~\ref{fig:quadrupole_lens}. The bore diameter is $2R=1.75~\mathrm{in}=4.45~\mathrm{cm}$, and the electrode length is $60~\mathrm{cm}$. For a good approximation of the electric quadrupole potential, the electrodes are chosen to be circular rods whose diameter is equal to the bore diameter~\cite{march2005quadrupole}.
%therefore we chose them to be the same for convenience.
The electrode length is $60~\mathrm{cm}$. Adjacent electrodes are held at spatially alternating potentials $\pm V$, with $V$ up to $30~\mathrm{kV}$. During high-voltage operation, a radiation monitor is mounted near the lens chamber to check for possible X-rays from electrode field emission or vacuum arcing~\cite{west2017underappreciated}. With the lens operating at $\pm30\rm kV$, the measured X-ray dose rate was approximately $0.08~\rm \mu Sv/h$, consistent with the ambient background level. The electrodes are mounted on Macor supports attached to stainless-steel rods. Flexible bellows at the chamber wall allow the rods to be moved by external three-axis translation stages at the upstream and downstream ends of the lens; common motion of the two stages sets the transverse position of the lens and differential motion sets its angle relative to the molecular beam axis while under vacuum.
%The rods are connected through vacuum bellows and are supported by external three-axis translation stages at the upstream and downstream ends of the lens, allowing the lens position and angle to be adjusted relative to the molecular beam axis.
The initial alignment was performed in air using an alignment laser and removable apertures.

\subsection{Detection and Lens Gain Measurement}
\label{sec:detection}
Downstream of the EQL, the molecules are detected by LIF in a chamber located $5.2~\mathrm{m}$ from the source. A rectangular aperture of width $20~\mathrm{mm}$ and height $50~\mathrm{mm}$ is positioned $25.4~\mathrm{mm}$ upstream of the probe laser. The aperture rejects molecules with large transverse velocities, whose Doppler shifts would move them far from resonance and reduce the scattered photons for fluorescence detection. It also defines the detection geometry, providing a well-defined solid angle as seen by photomultiplier tubes (PMTs).
%used in the molecular-number estimate.
Fluorescence is collected by two opposing PMTs mounted above and below the probe region, symmetrically about the molecular beam axis.

Detection is performed on the $R$-branch transition from the $J=2$ rotational level of the ground state to the highest hyperfine component of the $\tilde{J}'=3$ manifold in the $B^3\Pi_1$ excited state (Section \ref{sec:state_mapping}). This transition is \RtwoFfour, where primes
%$\tilde{F}_1'$ and $F'$
are used for excited-state quantum numbers,
%and $R$-branch indicates $\tilde{J}'=J+1$.
and $P$-, $Q$- and $R$-branch labels denote transitions with $\tilde{J}'=J-1$, $\tilde{J}'=J$ and $\tilde{J}'=J+1$, respectively.
%This $R$-branch transition
%It only addresses molecules in the highest hyperfine state, $F=3$, of the $J=2$ ground-state rotational manifold.
The transition wavelength is $271.75~\mathrm{nm}$.
%in the ultraviolet (UV) regime.
This ultraviolet (UV) probe light is generated by frequency quadrupling $1087~\mathrm{nm}$ seed laser output in two successive second-harmonic-generation stages, producing up to $50~\mathrm{mW}$ of the UV.
%at $271.75~\mathrm{nm}$.
The transition readily saturates and yields approximately two scattered photons per molecule before pumping into dark states~\cite{norrgard2017hyperfine,meijer2020lambda,grasdijk_rotational_cooling_2025}.

We determine the lens gain from
%the ratio of
LIF signals in the detection region.
%with the EQL energized and the EQL switched off, under otherwise identical state-preparation and detection conditions.
%Unless otherwise stated,
%By default, lens gain is determined from background-subtracted LIF signals
We normalize signals to an ablation-yield monitor that records $Q$-branch absorption immediately downstream from the beam-source cryogenic cell and provides a shot-to-shot measure of the molecular beam yield obtained from ablating a solid TlF target. Additional measurements where the probe position, beam size, and detuning were varied helped us characterize the molecular spatial and velocity distributions.

\subsection{State Mapping and Background Contributions}
\label{sec:state_mapping}

Each TlF eigenstate is labeled by the value of $J$ to which it connects adiabatically as the electric field is reduced to zero~\cite{Grasdijk_2021,CeNTREX_state_preparation_a}.
%In sufficiently large electric fields, and in states where $m_J=0$, the two nuclear spins
%%, , in certain regimes of $E$-field,
%are conveniently described by their total nuclear spin $\mathbf{I}_t=\mathbf{I}_1+\mathbf{I}_2$, giving singlet and triplet manifolds with $I_t=0$ and $I_t=1$, respectively.
In sufficiently large electric fields, the rotational angular momentum decouples from the nuclear spins, and $m_J$ becomes a good quantum number.
For $m_J=0$ states, the two nuclear spins are then conveniently described by the total nuclear spin $\mathbf{I}_t=\mathbf{I}_1+\mathbf{I}_2$, giving a singlet ($I_t=0$) and a triplet manifold ($I_t=1$). (The nuclear spin-rotation coupling averages to zero for $m_J=0$, leaving the nuclear spin-spin interaction.) For the stretched triplet states ($m_{I_t}=\pm1$) this coincides with states of definite $m_{I_1}$ and $m_{I_2}$, whereas the $m_{I_t}=0$ states are the symmetric (triplet) and antisymmetric (singlet) combinations.
%shown in Fig.~\ref{fig:state_evolution}.
%The good quantum numbers are $m_J$, $m_{I_1}$, and $m_{I_2}$
%%where $m_J$, $m_{I_1}$, and $m_{I_2}$
%which are the projections of $J$, $I_1$, and $I_2$,
%%along a quantization axis,
%respectively.

\begin{figure}
    \centering
\input{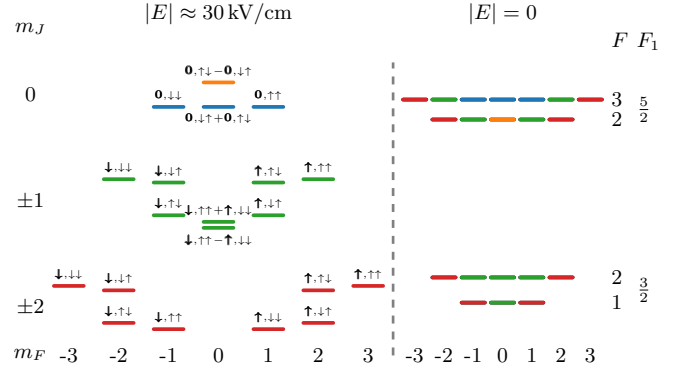}
    \caption{Mapping of the $J=2$ manifold as the electric field magnitude is reduced from $|E|\approx 30~\rm kV/cm$ to zero. Both sides are arranged in columns of $m_F$, which is conserved throughout. The high-field states (left) are labeled by $m_J$ and by their nuclear-spin components $m_{I_1},m_{I_2}$ as, for example, $\spinpair{\uparrow}{\upd}$: the first, bold symbol denotes the sign of $m_J$, a bold $0$ for $m_J=0$, while the latter two arrows denote $m_{I_i}=\pm 1/2$. The zero-field states (right) are labeled in the coupled basis $\ket{F_1,F,m_F}$, whose $m_F$ sublevels are degenerate. Colors indicate the correlations: green and red denote $|m_J|=1$ and $|m_J|=2$, respectively, blue and orange the $m_J=0$ states with total nuclear spin $I_t=1$ and $I_t=0$, respectively. The zero-field side is drawn to scale. On the high-field side, splittings within each $m_J$ group are to scale, except where too small to render. Different magnification factors are used for different $m_J$ groups, and the groups themselves are drawn equally spaced.}
    \label{fig:state_evolution}
\end{figure}

For the lensing measurements reported here, the rotational-cooling laser was operated on the \PtwoFtwo{} transition rather than on the \PtwoFone{} that will be used under ordinary operating conditions for CeNTREX. This prepares molecules
%and detects population
in
%the nuclear-spin triplet manifold
($I_t=1$) of the $J=0$ state to take advantage of a favorable detection scheme, while the \CENTREX{} Schiff-moment measurement will operate on
%in the singlet manifold
($I_t=0$).
%the present lensing measurements are performed in the triplet manifold because it provides a favorable detection scheme.
Since the $m_J=0$ states have nearly identical Stark shifts over the relevant electric field range, their lensing behavior is expected to be the same.

Detection is performed at nominally zero electric field.
%where the states are described in the coupled hyperfine basis.
Figure~\ref{fig:state_evolution} shows the adiabatic mapping between the high-field
%uncoupled basis and the zero-field coupled basis
and zero-field bases in the $J=2$ manifold. The ($m_J=0,I_t=1$) triplet (blue states in Fig.~\ref{fig:state_evolution}) maps to
$\ket{F_1=5/2,F=3,m_F=0,\pm1}$,
which are addressed by the detection laser. The ($m_J=0,I_t=0$) singlet (orange state in Fig.~\ref{fig:state_evolution}) maps to
$\ket{F_1=5/2,F=2,m_F=0}$, which
%selection rules dictate
cannot be excited to the final state $F'=4$ by the probe laser.

The choice of detection transition is important for suppressing coherent dark states. In the $X^1\Sigma^+$ ground state the hyperfine splittings are $\sim100~\mathrm{kHz}$, much smaller than the natural linewidth of the $B^3\Pi_1$ state, $\Gamma_B=1.6~\mathrm{MHz}$. Thus, multiple ground-state hyperfine components are simultaneously addressed by a single laser frequency. In contrast, the excited-state hyperfine structure is resolved, with splittings ranging from $\sim100~\mathrm{MHz}$ to $\sim13~\mathrm{GHz}$.
%Driving an inappropriate excited-state hyperfine component can therefore produce coherent dark states and reduce the fluorescence yield.
To avoid producing coherent dark states and reducing the fluorescence yield, we drive the transition from $\ket{J=2,F_1=5/2,F=3}$ to the highest hyperfine component ($F'=4$) of the $\tilde{J}'=3$ manifold. This connects seven ground states to nine excited states and prevents dark states.

Residual population in other $J=2$ states, left over from imperfect rotational cooling, can contribute a small background to the measured lens gain. The $m_J=\pm1$ states exhibit only weak lensing, while the $m_J=\pm2$ states are defocused by the lens.
%so their relative contribution is reduced when the EQL is energized.
%In our simulations, we ignore their contributions.
%Rotational cooling removes most of the initial population from the $J=2$ manifold, although a small residual population remains.
Based on the measured rotational-cooling efficiency~\cite{grasdijk_rotational_cooling_2025}, the residual background is expected to be at the percent level at zero lens voltage and smaller at nonzero voltage; we ignore it in our simulations.  We note that the lens also acts as an $m_J=0$ state filter.

\subsection{Trajectory Simulations}
\label{subs:trajectory_simulations}

We used numerical trajectory simulations to model the lens behavior and to test this model against the measurements.
In the simulations, transverse molecular motion is governed by
%\begin{equation}
%\label{eq:EOM}
%    m\frac{d^2\mathbf r_\perp}{dt^2}
%    =
%    -\nabla_\perp U\!(\mathbf r_\perp),
%\end{equation}
\begin{align}
\label{eq:EOM}
  m\,\frac{d^{2}X}{dt^{2}} &= -\frac{\partial (\Delta E_{S})}{\partial X}, &
  m\,\frac{d^{2}Y}{dt^{2}} &= -\frac{\partial (\Delta E_{S})}{\partial Y}-mg,
\end{align}
where $\Delta E_{S}$ is the exact Stark shift of the
lensing state [Fig.~\ref{fig:quadrupole_lens}(a)] and $g$ is the gravitational acceleration.
%and $g$ is the acceleration due to gravity.
%where $U$ is the Stark shift of the lensing state (shown in Fig. \ref{fig:quadrupole_lens}(a)) calculated from the full hyperfine Hamiltonian, and $\mathbf r_\perp$ is the transverse position.
%and $m$ is the mass of TlF molecule.
In contrast to the model of Section~\ref{subs:lens_model}, the simulations integrate the full two-dimensional transverse motion.
Within the lens region, we used the ideal quadrupole electric field described by Eq.~(\ref{eq:equad}). We also calculated the full three-dimensional electric field distribution of the lens using COMSOL software including fringe field effects at the entrance and exit of the lens.
%including both transverse and longitudinal dependence.
The resulting trajectory simulations using this field distribution showed negligible differences from those obtained using the ideal quadrupole field, so the ideal-field approximation was used for the results presented here. Equation~(\ref{eq:EOM}) was numerically integrated using the Runge-Kutta-Fehlberg method (RK45) to obtain the position and velocity of each molecule after propagation through the lens region. Outside the lens region, the molecules were assumed to move ballistically. The initial trajectories were sampled from the measured source position distribution (Section~\ref{subs:source_dist}) together with the measured source velocity distribution.

\section{Results}
\label{sec:results}

First, we present the measurement of the molecular source position distribution. Then we characterize the EQL using three observables: the lens gain, the transverse velocity distribution, and the transverse spatial distribution at the detection region. The lens gain directly quantifies the increase in detected molecular flux, while the velocity and spatial distributions provide independent tests of our trajectory simulation model and determine the beam properties needed to design the downstream state-preparation regions.

\subsection{Source Distribution Characterization}
\label{subs:source_dist}

The transverse position distribution of molecules exiting the cryogenic beam source was characterized using a one-dimensional pinhole-camera geometry. A narrow horizontal slit of \(25.4~\mathrm{mm}\) width and \(2~\mathrm{mm}\) height was placed inside the RC chamber \(51~\mathrm{cm}\) downstream from the source, effectively serving as a pinhole. The detection chamber was located \(151~\mathrm{cm}\) downstream from the beam source, giving an image-to-object distance ratio of \(\sim2{:}1\) and a geometric magnification of \(\sim2\). The detection laser had a $1/e^2$ diameter of $3.2~\rm mm$ along the vertical direction. The vertical position of the laser beam was scanned using a periscope mounted on a translation stage, and LIF signal was recorded as a function of laser beam height to obtain a one-dimensional magnified projection of the source distribution.
%magnified by approximately \(2\). 

\begin{figure}
    \centering
    \includegraphics[width=8cm]{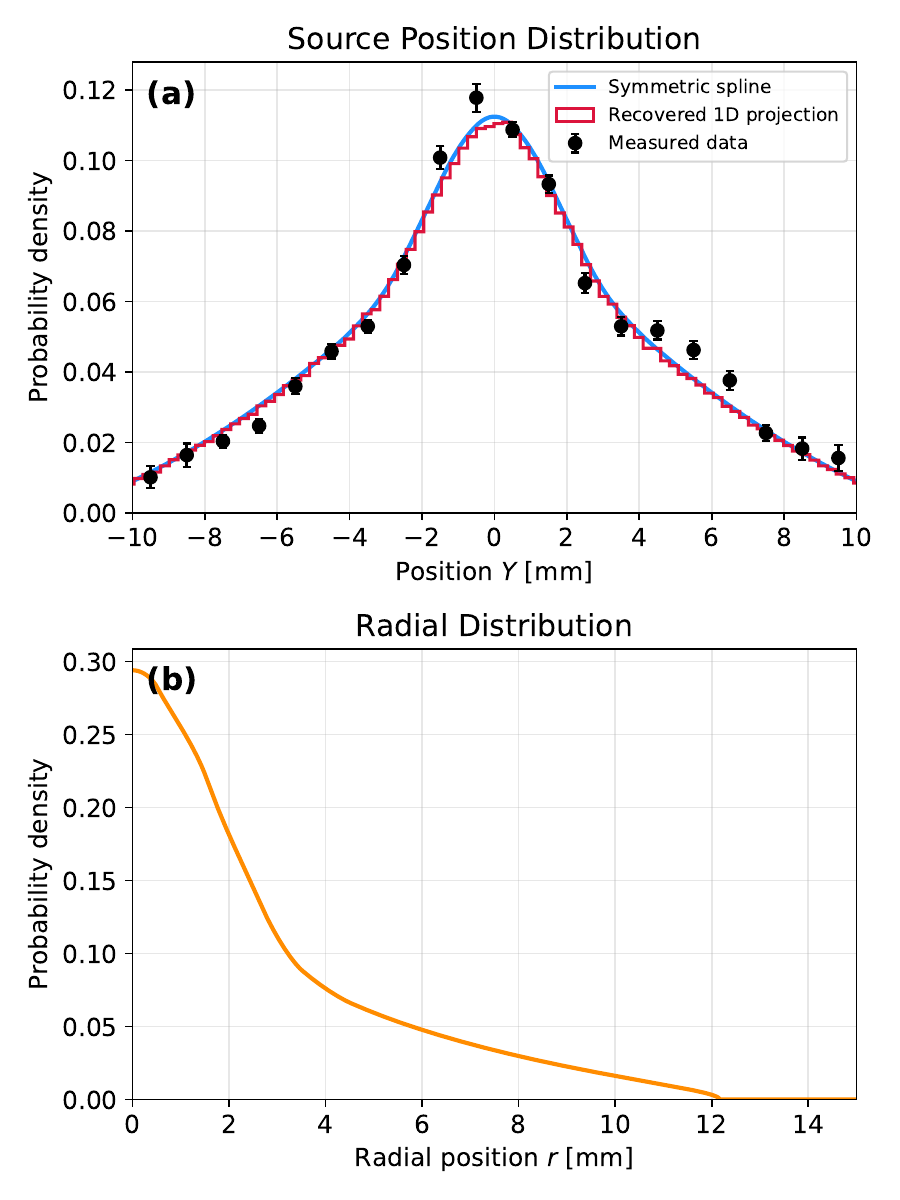}
    \caption{(a) Measured one-dimensional projection of the molecular source position distribution, normalized by its integral (black points); the symmetrized spline interpolation used as input to the trajectory simulations (blue line); and the one-dimensional projection obtained by sampling the reconstructed radial distribution (red line). (b) The radial position distribution recovered by the inverse Abel transformation.  These measurements were used as input for the trajectory simulations.}
    \label{fig:source_dist}
\end{figure}
This projection is shown in Fig.~\ref{fig:source_dist}(a). Assuming approximate azimuthal symmetry of the source emission about
the beam axis, which is consistent with the cylindrical symmetry of the source cell aperture, the two sides of the measured projection should be identical within measurement noise~\cite{hickstein2019}. We therefore symmetrized the projection about its centroid and fitted the averaged half-profile with a cubic spline [Fig.~\ref{fig:source_dist}(a), blue curve] before applying the inverse Abel transform to recover the radial
%two-dimensional
source distribution in Fig.~\ref{fig:source_dist}(b)~\cite{Dasch92}. As a consistency check, we sampled this radial distribution and projected onto a single Cartesian axis, yielding the red curve in Fig.~\ref{fig:source_dist}(a). This is the initial transverse position distribution used in the trajectory simulations of Section \ref{subs:trajectory_simulations}. After correcting for the magnification, the reconstructed distribution has a full width at half maximum of \(\sim5.4~\mathrm{mm}\). 
%Trajectory simulations comparing the assumed source distribution with the distribution that would be inferred from the slit-based measurement indicate
We estimate that the finite slit height contributes a systematic error of \(\lesssim 0.2~\mathrm{mm}\) to the source size. This error is small compared to other systematic effects that arise in lens characterization and therefore is not included further.
%through the simulations.

\subsection{Lens Gain}

The lens gain at the center is defined as the ratio of the background-subtracted, yield-normalized LIF signal with the EQL energized to that with the EQL turned off. For this measurement, the detection laser had
%was expanded vertically to
a $1/e^2$ diameter of $8.4~\mathrm{mm}$ vertically and $2.6~\rm mm$ horizontally, with a power of $24~\mathrm{mW}$. The corresponding power-broadened linewidth is $16.7~\mathrm{MHz}$, sufficient to saturate molecules passing near the center of the detection area. The detection area was placed 5.2 m from the source and 3.8 m from the center of the lens.

We determined the forward velocity distribution of the molecular beam to be approximately Gaussian, with its mean $\bar{v}_Z$ drifting slowly with depletion of the ablation target. To reduce sensitivity to this drift, the lens on/off gain was measured in an interleaved sequence. For most lens voltage settings, the gain was obtained from two interleaved and averaged on/off measurements; near the maximum gain at $26.4~\mathrm{kV}$, three measurements were made.
%The plotted gain at each voltage is the mean of the repeated measurements at that voltage, and the uncertainty is estimated from the standard error of the mean of the corresponding spot-averaged signals.
For the voltage-on points collected mid-measurement,
%additional
interleaving effectively doubled the lens-off sampling since
%to 64 spots.
these points were bracketed by lens-off shots on both sides.
The forward velocity distribution was extracted \textit{in situ} from the time-of-flight traces, and the measured gains were rescaled to a nominal mean forward velocity of $195~\mathrm{m/s}$ using a velocity-dependent correction factor obtained from our simulations. This factor takes a value ranging from 0.95 to 1.1.

\begin{figure}
    \centering
    \includegraphics[width=\linewidth]{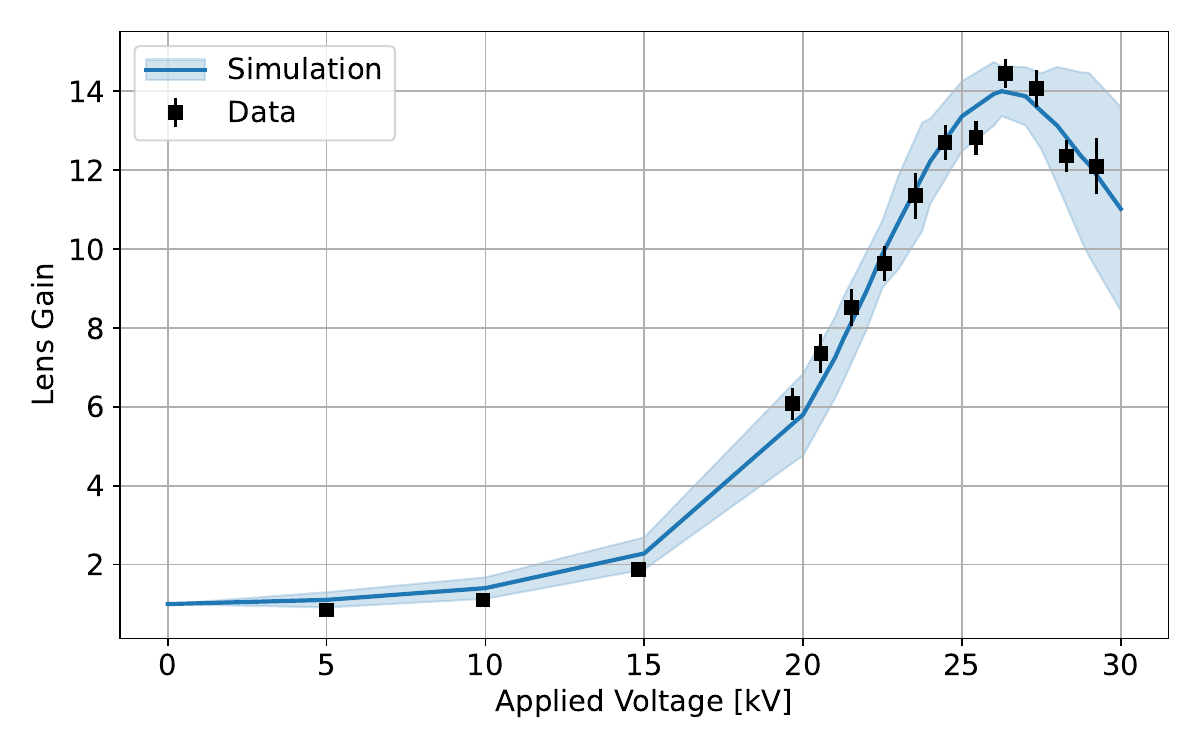}
    \caption{Lens gain as a function of applied EQL voltage. The points show the measured, background-subtracted, yield-normalized LIF gain, rescaled to a nominal mean forward velocity of $195~\mathrm{m/s}$. Error bars represent the standard error of the mean.
    %from repeated measurements.
    The solid curve shows the trajectory simulation using the same nominal velocity distribution; the shaded band indicates the effect of measured shot-to-shot fluctuations in the mean forward velocity.}
    \label{fig:lens_gain}
\end{figure}

\begin{figure}[htbp]
    \centering
    \includegraphics[width=8.5cm,width=\linewidth]{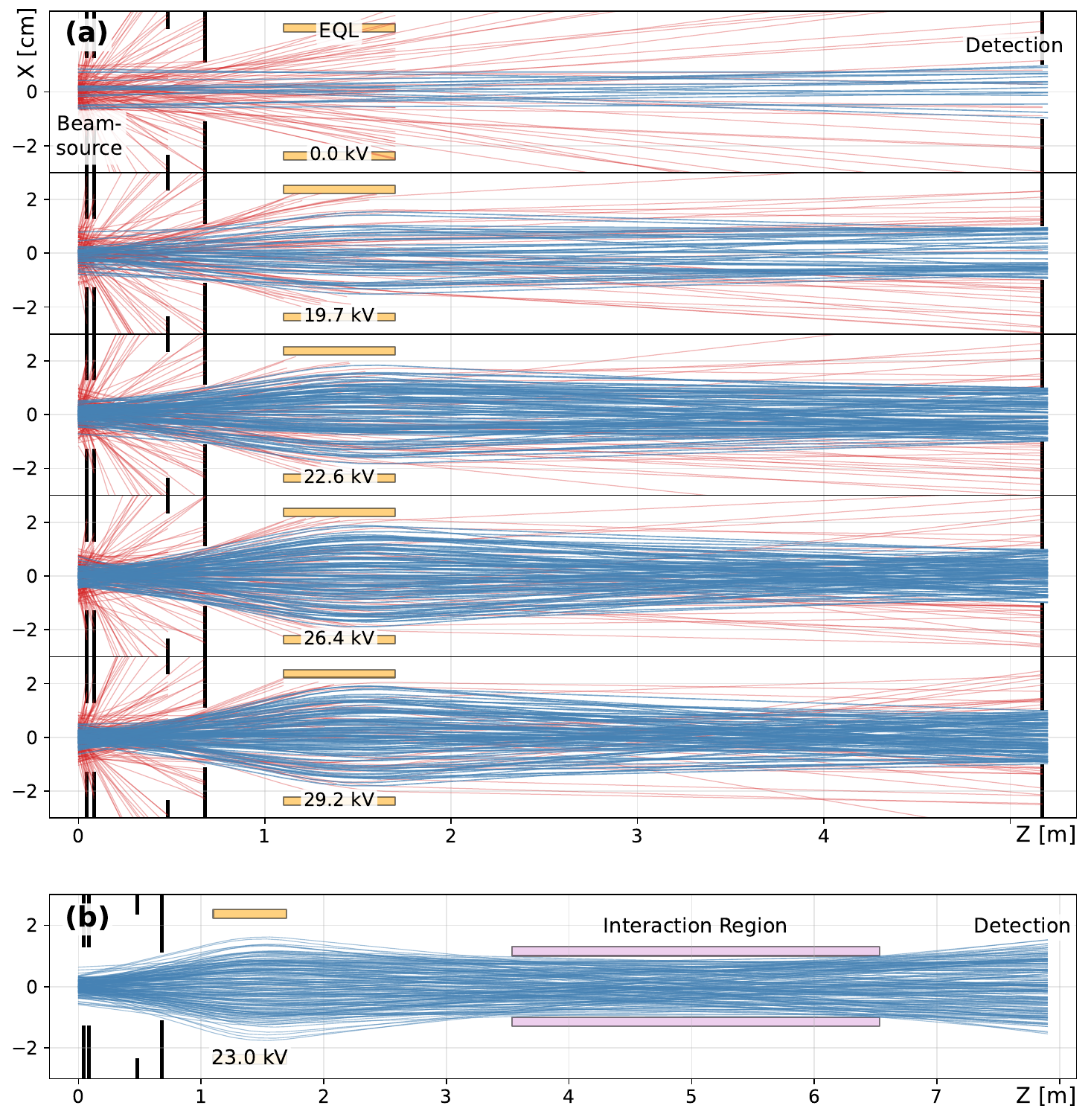}
    \caption{(a) Trajectory simulations showing lens focusing at various voltages for the current experimental setup. Blue lines represent trajectories of observed molecules, while red lines represents trajectories of molecules that are rejected by apertures. The lens focus moves from right to left as the voltage increases, and at $26.4~\rm kV$ it coincides with the detection region.
    %At lower voltages, the focusing effect of the lens increases as the voltage increases; at $26.4~\rm kV$, the focal point of the lens shifts near the detection region, resulting highest measured gain; at higher voltages, the lens starts to overfocus and the focusing point shifts upstream of detection region.
    (b) Trajectory simulations for the full $7.9\;\rm m$ CeNTREX beamline, at the optimal voltage of $23\;\rm kV$.}
    \label{fig:trajectory_simulations}
\end{figure}

Figure~\ref{fig:lens_gain} shows the measured lens gain as a function of EQL voltage. The gain increases with voltage, reaches a maximum of 14.4(4) at $26.4~\mathrm{kV}$, and decreases at higher voltages. The ray transfer model of Section \ref{subs:lens_model} predicts approximate collimation near \(26~\mathrm{kV}\), where the focal region overlaps the detection volume.
%while the maximum detected gain occurs near \(27~\mathrm{kV}\), where the focal region overlaps the detection volume.
The observed voltage dependence is consistent with the expected focusing behavior: increasing voltage increases the focusing strength, moving the focal plane upstream
%downstream
until it approaches the detection region. At higher voltages,
%the focal plane moves upstream,
overfocusing occurs and the beam expands before detection. This dynamic is illustrated by Fig.~\ref{fig:trajectory_simulations}. 

rigwhat Trajectory simulations (Section~\ref{subs:trajectory_simulations}) were used to calculate the lens gain at its center, for molecules with nominally zero transverse velocity. The simulations utilized the measured source position distribution, the measured transverse velocity distribution, and the nominal forward velocity distribution used in the rescaling procedure. The solid curve in Fig.~\ref{fig:lens_gain} shows the simulated gain at the nominal mean forward velocity, while the shaded band includes the effect of the measured shot-to-shot mean forward-velocity fluctuations over $\sim3~\rm min$ measurement period when the lens is turned on. The simulated gain at the center is $14.0(6)$ while the simulated gain integrated over the entire molecular beam is $13.8(4)$ at optimal voltage. The measured gain is fully consistent with the trajectory simulations combined with the optical Bloch equation (OBE) LIF detection model. 

\subsection{Transverse Velocity Distribution}
\label{subs:TransVelDistr}

\begin{figure*}
    \centering
    \includegraphics[width=18cm]{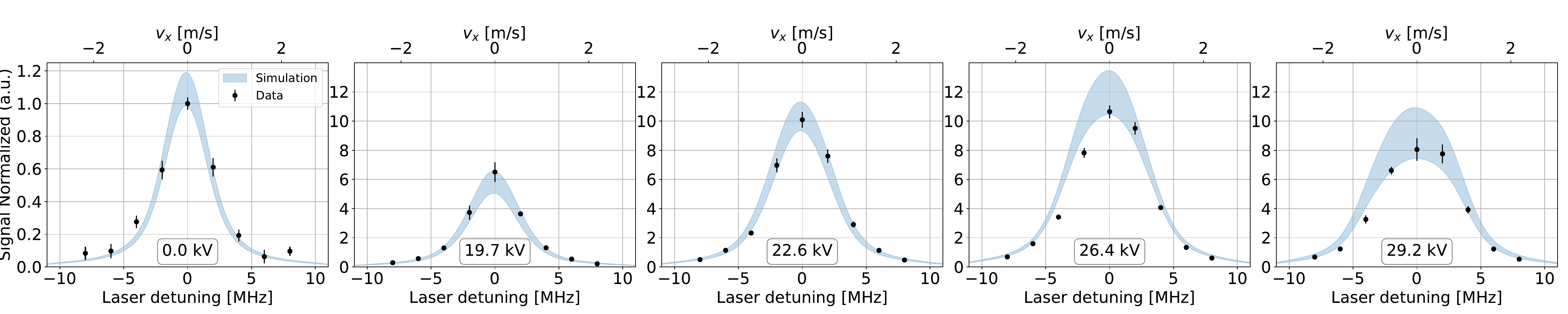}
    \caption{Transverse (horizontal) Doppler spectra for several EQL voltages. A transverse velocity of \(1~\mathrm{m/s}\) corresponds to \(3.68~\mathrm{MHz}\) of detuning at \(271.75~\mathrm{nm}\).
    %(all detunings are quoted at the UV frequency).
    The data was normalized to unity at zero detuning and $0~\rm kV$. Error bars show the standard error of the mean. The shaded simulation bands include the measured forward-velocity fluctuations. Note the different vertical scale in the leftmost plot.
    %change in $y$-scales between different plots.
    }
    \label{fig:transverse_v_distribution}
\end{figure*}

The transverse velocity $v_X$ distribution is characterized by scanning the detection laser detuning and measuring the Doppler spectra. Figure \ref{fig:transverse_v_distribution} shows the Doppler spectra at various lens voltages.
The detection laser was aligned to the $X$-axis (Fig. \ref{fig:experiment_schematic}) of the beamline, with $0.5~\rm mW$ power to minimize power broadening. Each data point is from 8 distinct laser-ablation locations on the TlF target, with 20 shots per location.
%with error bars showing the standard error of the mean over spot-averaged signals.
%The experimental data was normalized to 1 when the lens was off and laser with zero detuning.

The measured Doppler FWHM increases from $4.8\,\mathrm{MHz}$ ($1.3\,\mathrm{m/s}$) at $0\,\mathrm{kV}$ to $6.4\,\mathrm{MHz}$ ($1.8\,\mathrm{m/s}$) at $26.4\,\mathrm{kV}$, and broadens to $7.2\,\mathrm{MHz}$ ($2.0\,\mathrm{m/s}$) at $29.2\,\mathrm{kV}$. As the voltage increases, the signal grows for all transverse velocities while the spread of the transverse velocity also increases. Above the optimum voltage, the Doppler width broadens rapidly due to overfocusing, and the signal starts to decrease. For an ideal monochromatic beam and a point source, focusing would reduce the transverse velocity spread to zero at the detection region. However, the longitudinal velocity spread produces chromatic aberration, causing the observed Doppler width to increase gradually with voltage up to the optimum voltage. Assuming approximate cylindrical symmetry of the source and lens geometry, the transverse velocity distribution along $Y$ is expected to follow the same trend. At the optimum \(26.4~\mathrm{kV}\) voltage, trajectory simulations indicate that the lens accepts and transports an upstream transverse-velocity distribution with an effective FWHM of $3.8~\mathrm{m/s}$ to the detection region. This acceptance is within the effective range of the upstream RC region~\cite{grasdijk_rotational_cooling_2025}. Moreover, the $1.8\,\mathrm{m/s}$ transverse velocity FWHM measured downstream is compatible with our subsequent interaction regions planned for the \CENTREX{} beamline~\cite{Grasdijk_2021}. 

The simulated Doppler spectra are obtained by convolving the lineshapes derived from OBEs, the transverse velocity distributions from trajectory simulations, and the seed laser lineshape (with a linewidth of $\sim300~\rm kHz$ prior to frequency quadrupling, and $\sim1.2~\rm MHz$ in UV, determined by beat measurement of two seed lasers). To compare the simulations with measurements, a single amplitude scaling parameter was derived by fitting the 0 kV normalized data to the simulation; the same normalization factor was then applied to all subsequent simulations. The simulated spectra shown in Fig. \ref{fig:transverse_v_distribution} reproduce both the increase in integrated signal and the voltage-dependent broadening of the Doppler profile.
%The shaded bands reflect the measured shot-to-shot variation in the forward velocity. 

\subsection{Transverse Position Distribution}

\begin{figure*}
    \centering
    \includegraphics[width=18cm]{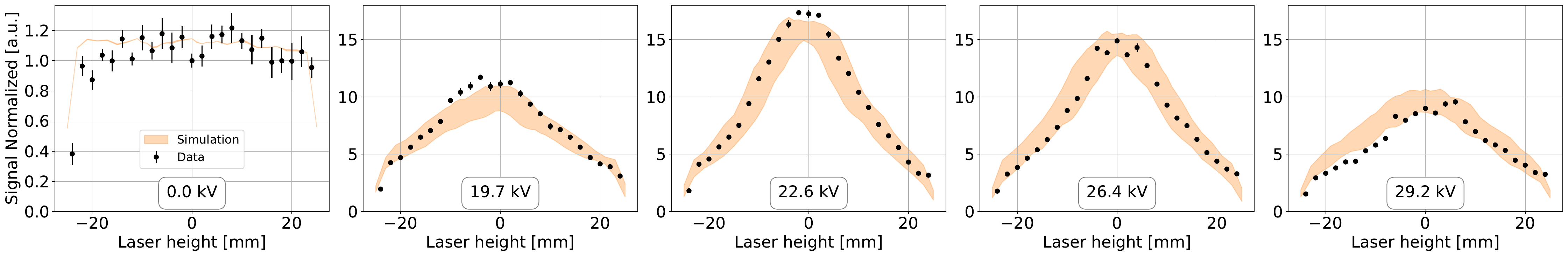}
    \caption{Transverse (vertical) molecular beam profiles for several EQL voltages.
    %The detection laser, with a \(1/e^2\) diameter of \(3.2~\mathrm{mm}\), was scanned in the vertical direction.
    The data was normalized to unity at zero height offset and $0~\rm kV$. Error bars show the standard error of the mean. The shaded simulation bands include the measured forward-velocity fluctuations and the estimated lens alignment uncertainty. Note the different vertical scale in the leftmost plot.}
    \label{fig:transverse_y_distribution}
\end{figure*}

The transverse position
%$Y$
distribution was measured by scanning the vertical ($Y$) position of the detection laser while maintaining a tighter vertical
%The detection laser power was set to 24 mW and with a smaller vertical
$1/e^2$ beam diameter of 3.2 mm for better resolution. Figure \ref{fig:transverse_y_distribution} shows the measured transverse spatial profiles of the molecules at various lens voltages. With the lens turned off, the vertical position distribution is relatively flat due to the large initial transverse velocity spread at the source.
%and the large transverse position spread as the molecules exit the cell and propagate along the beamline.
As the lens voltage increases, the distribution narrows
%and becomes strongly peaked
due to focusing. However, as the voltage increases further, the molecular beam becomes overfocused, resulting in a reduced overall signal amplitude and a flatter spatial distribution.
%near the center.
Assuming cylindrical symmetry, we expect the horizontal ($X$) spatial distribution to follow the same trend, aside from the slight vertical gravitational drop.

For each probe height, the signals collected from the upper and lower PMTs were corrected for their height-dependent collection solid angles before averaging to obtain the normalized molecular signals. The opposing-PMT geometry reduces sensitivity to height-dependent collection efficiency variations. Each data point in Fig.~\ref{fig:transverse_y_distribution} is from 8 distinct laser-ablation locations, with 10 shots per location.
%error bars show the standard error of the mean over spot-averaged signals. The experimental data was normalized to 1 when the lens was off and laser at zero height offset.

Trajectory simulations were used to compute the expected molecular position distributions.
%with error bands derived from the measured forward velocity fluctuations.
The simulated profiles take into account
%include convolution with
the measured intensity profile of the detection laser. The single-parameter normalization described in Section \ref{subs:TransVelDistr} was applied here as well.
%To compare the simulations with measurements, a single amplitude scaling parameter was derived by fitting the 0 kV normalized data to the simulation; the same normalization factor was then applied to all subsequent simulations.

Our simulations predict that the center of the molecular distribution should drop by $\sim2$ mm relative to the reference height due to gravity. However, the observed distribution is centered near 0 mm. We attribute this to the alignment accuracy of the lens. A lens electrode tilt of \(\sim5\times10^{-4}~\mathrm{rad}\), which is within our alignment tolerance, would compensate the gravitational drop. The effect of the tilt was incorporated into the error band calculations.

\section{Conclusion}
\label{sec:conclusion}
We have designed, implemented, and characterized an electrostatic quadrupole lens to focus a cryogenic molecular beam of $^{205}$TlF. By exploiting the quadratic Stark effect of the $|J=2, m_J=0\rangle$ levels in the $X^1\Sigma^+$ ground state, the quadrupole electrode configuration generates a radially linear electric field resulting in a harmonic restoring force. This mechanism serves to transversely focus the molecular beam, analogously to a thick optical lens, with an effective acceptance of transverse velocities within \(\pm 3.8~\mathrm{m/s}\). This should be compared to the aperture-limited geometric acceptance of only $\pm 0.4$~m/s.

We used laser-induced fluorescence to measure a maximum signal enhancement of $14.4(4)$ at an optimal applied voltage of $\pm26.4~\rm kV$, where the focal point of the lens overlaps with the detection position. The experimentally determined lens gain, transverse Doppler spectra, and transverse spatial profiles are all reproduced by trajectory simulations using independently measured molecular and laser beam parameters, validating our model of the lens dynamics and confirming the adiabaticity of molecular state propagation within the lens.

The full \CENTREX{} beamline is \(7.9~\mathrm{m}\) long. In this case, the trajectory simulations project the lens gain at center to be $15.3(5)$. The simulated gain integrated over the entire molecular beam is $16.5(8)$, closely comparable to the ThO hexapole result with linear Stark shifts \cite{Wu_2022} and translating directly into a statistical sensitivity for the $^{205}\rm Tl$ Schiff moment measurement improved by a factor of $4.1(1)$ in the shot-noise limit.
%at the location of the final interaction region, where the longer distance from the lens further enhances the focusing benefit.
The EQL works in concert with the upstream rotational-hyperfine cooling and microwave adiabatic-passage stages of the \CENTREX{} state-preparation protocol, and the demonstrated enhancement of the detected molecular signal translates directly into improved statistical sensitivity for the planned Schiff moment measurement. Ultimately, these results establish the EQL as a vital component for maximizing molecular flux and increasing sensitivity in precision measurements that utilize cryogenic polar-molecule beams.  

\begin{acknowledgments}
We are grateful for support from the John Templeton Foundation; the Heising-Simons Foundation Grant 2022-3842; a NIST Precision
Measurement Grant; NSF-MRI Grants No. PHY-1827906 and No. PHY-1828097; NSF Grant No.
PHY-2110420; and the Department of Energy (DOE), Office
of Science, Office of Nuclear Physics, under Contract No.
DEAC02-06CH11357 and Grant No. DE-SC0024667.

% We are grateful for support from the NSF Grant No.
% PHY-2110420, Heising-Simons Foundation Grant 2022-3842.
\end{acknowledgments}

% \appendix

% \input{appendix/appendix}

% \nocite{*}

\bibliography{references}% Produces the bibliography via BibTeX.

\end{document}